\documentstyle[aps,prc,epsfig]{revtex}
\draft

\newcommand{\ba}{\begin{eqnarray}}
\newcommand{\ea}{\end{eqnarray}}
\newcommand{\ii}{\'{\i}}

\begin{document}

\title{Single-Particle Transfer and Nuclear SUperSYmmetry. \\ 
Pick-Up and Stripping with SUSY}

\author{J. Barea$^{1)}$, R. Bijker$^{2)}$, A. Frank$^{2,3)}$ 
and G. Loyola$^{2)}$}

\address{
$^{1)}$ Departamento de F\ii sica At\'omica, Molecular y Nuclear, 
Facultad de F\ii sica, \\ 
Universidad de Sevilla, Apdo. 1065, 41080 Sevilla, Spain \\
$^{2)}$ Instituto de Ciencias Nucleares, 
Universidad Nacional Aut\'onoma de M\'exico, \\
Apartado Postal 70-543, M\'exico, D.F., M\'exico \\
$^{3)}$Centro de Ciencias F\ii sicas, 
Universidad Nacional Aut\'onoma de M\'exico, \\
Apartado Postal 139-B, Cuernavaca, Morelos, M\'exico}

\maketitle

\begin{abstract}
Transfer reactions constitute a stringent test for nuclear 
supersymmetry, a theory that simultaneously describes neighboring nuclei
with bosonic and fermionic character. We construct and analytically
evaluate one-nucleon transfer matrix elements between supersymmetric
partners with the $U(6/4)$ case as an example, and stress the need for a
careful treatment of bosonic and fermionic operators in the construction
of mixed tensor operators.
\end{abstract}

\section{Introduction}

Recently the supersymmetric classification of nuclear levels in the Pt 
and Au isotopes has been re-examined by taking advantage of the significant
improvements in experimental capabilities developed in the last decade.
High resolution transfer experiments with protons and polarized deuterons 
have led to strong evidence for the existence of supersymmetry (SUSY) in 
atomic nuclei. The experiments include high resolution transfer experiments 
to $^{196}$Au at TU/LMU M\"unchen \cite{metz,pt195}, and in-beam gamma ray 
and conversion electron spectroscopy following the reactions 
$^{196}$Pt$(d,2n)$ and $^{196}$Pt$(p,n)$ at the cyclotrons of the PSI and 
Bonn \cite{au196}. These studies have achieved an improved classification 
of states in $^{195}$Pt and $^{196}$Au which give further support to the 
original ideas \cite{baha,sun,magic} and extend and refine previous 
experimental work \cite{mauthofer,jolie,rotbard} in this research area.

In the context of the Interacting Boson Model \cite{IBM} and its extensions
\cite{IBFM}, Iachello and his co-workers proposed that Bose-Fermi symmetries 
$U(6) \otimes U(\Omega)$ can be embedded into a graded Lie algebra 
$U(6/\Omega)$ in 
order to unify even-even and even-odd nuclei \cite{FI,susy} in a 
supersymmetric framework. The supersymmetric irreducible representation 
$[{\cal N}\}$ encompasses the even-even nucleus with ${\cal N}$ bosons and 
the odd-even nucleus with ${\cal N}-1$ bosons and the odd fermion. This idea 
was later extended by Van Isacker and his co-workers to the case where
neutron and proton bosons are distinguished \cite{magic}, predicting in
this way a correlation  among quartets of nuclei, the even-even partner
with $N_{\nu} + N_{\pi}$ bosons, the odd-proton and odd-neutron neighbors
with $N_{\nu} + N_{\pi} - 1$ bosons, and the odd-odd nucleus with
$N_{\nu} + N_{\pi} - 2$ bosons.

While this simultaneous description and classification has been shown to 
be fulfilled to a good approximation for the quartet of nuclei $^{194}$Pt, 
$^{195}$Au, $^{195}$Pt and $^{196}$Au, there are important predictions 
still not fully verified by experiments. These tests involve the transfer 
reaction intensities among the supersymmetric partners. 
In this Brief Report we discuss the example of proton transfer between 
the SUSY partners $^{194}$Pt and $^{195}$Au 
(e.g. through the $(\vec d, n)$ stripping reaction) in the $U(6/4)$ scheme.  
We emphasize that special care has to be taken in the theoretical analysis 
because of phase ambiguities in the $SO(6)$ classification \cite{phases} 
of boson and fermion tensor operators. 

\section{The $U(6/4)$ supersymmetry}

First, we briefly review the $U(6/4)$ supersymmetry. For the sake of 
simplicity, just as in the original papers \cite{FI,susy,spin6}, we do 
not distinguish between neutron and proton 
bosons. The simultaneous description of $^{194}$Pt and $^{195}$Au nuclei in 
the SUSY scheme assumes that the odd-proton hole in $^{195}$Au can only 
occupy the $2d_{3/2}$ orbit. The $U(6/4)$ dynamical supersymmetry
involves the group chain
\ba
\left| \begin{array}{ccccccccc} U(6/4) &\supset& 
U^B(6) &\otimes& U^F(4) &\supset& SO^B(6) &\otimes& SU^F(4) \\
\, [{\cal N}\} && [N] && \{1^M\} && (\Sigma,0,0) && \end{array} \right. 
\nonumber \\
\mbox{} 
\nonumber \\
\hspace{1cm} \left. \begin{array}{cccccc}
\supset& Spin(6) &\supset& Spin(5) &\supset& Spin(3) \\
& (\sigma_1,\sigma_2,\sigma_3) && (\tau_1,\tau_2) && J \end{array} \right> ~. 
\ea
The corresponding Hamiltonian, neglecting terms that contribute only to
the binding energy, is of the form
\ba
H &=& A \, C_{2SO^B(6)} + B \, C_{2Spin(6)} 
+ C \, C_{2Spin(5)} + D \, C_{2Spin(3)} ~, 
\ea
which leads to an analytic form for the energy as a function of the
quantum numbers 
\ba
E &=& A \, \Sigma(\Sigma+4) + B \, \left[ \sigma_1(\sigma_1+4) 
+ \sigma_2(\sigma_2+2) + \sigma_3^2 \right]
\nonumber\\
&& + C \, \left[ \tau_1(\tau_1+3) + \tau_2(\tau_2+1) \right] 
+ D \, J(J+1) ~.
\ea
The coefficients $A$, $B$, $C$ and $D$ are determined in a 
simultaneous fit of the excitation energies of even-even and odd-even 
nuclei which belong to the same supermultiplet. In addition to energies, in 
the case of a 
dynamical supersymmetry closed expressions can be derived for other 
observables as well, such as electromagnetic intensities and single-particle 
transfer reaction strengths. In the next section we concentrate on the 
latter, and in particular on the transfer between the even and odd members 
of the supermultiplet. 

\section{Single-particle transfer}

The single-particle transfer operator that is commonly used in the 
Interacting Boson-Fermion Model (IBFM), 
has been derived semi-microscopically in the seniority scheme \cite{olaf}. 
Although strictly speaking this derivation is only valid in the vibrational 
regime, it has been used for deformed nuclei as well. An alternative method 
is based on symmetry considerations. It consists in expressing the 
single-particle transfer operator in terms of tensor operators 
under the subgroups that appear in the group chain of a dynamical 
(super)symmetry \cite{spin6,BI}. 

For single-particle transfer between different supermultiplets of 
$U(6/4)$ characterized by $[{\cal N}\}$ and $[{\cal N}+1\}$, the 
transfer operator, to lowest order, is given by $a^{\dagger}_{3/2,m}$ 
which by construction transforms as a $SU(4)$ tensor operator 
\ba
P^{\dagger}_{\frac{3}{2},m} &=& \alpha \, a^{\dagger}_{\frac{3}{2},m} \;=\; 
\alpha \, T^{(\frac{1}{2},\frac{1}{2},\frac{1}{2})}
_{(\frac{1}{2},\frac{1}{2}),\frac{3}{2},m} ~. 
\label{pa1}
\ea
Here the upper indices denote $(\sigma_{1},\sigma_{2},\sigma_{3})$, which 
specify the tensorial properties under $Spin(6)$ which is isomorphic to 
$SU(4)$ and $SO(6)$ \cite{susy,spin6}. Similarly, 
the lower indices represent $(\tau_{1},\tau_{2})$, $J$ and $M$, 
which define the transformation properties under $Spin(5)$, 
$Spin(3)$ and $Spin(2)$, respectively.    
For the inverse process we have
\ba
\tilde{P}_{\frac{3}{2},m} &=& \alpha \, \tilde{a}_{\frac{3}{2},m} \;=\; 
\alpha \, T^{(\frac{1}{2},\frac{1}{2},-\frac{1}{2})}
_{(\frac{1}{2},\frac{1}{2}),\frac{3}{2},m} ~,
\label{pa2} 
\ea
with $\tilde{a}_{j,m}=(-1)^{j-m} a_{j,-m}$~.

The most stringent test of supersymmetries, however, is provided 
by the single-particle transfer between different members of the 
same supermultiplet. The transfer operator in this case involves the 
interchange of a boson and a fermion operator 
\ba
P^{\dagger}_{\frac{3}{2},m} &=& \sum_{l=0,2} \alpha_l \, 
\left( b^{\dagger}_l \times \tilde{a}_{\frac{3}{2}} \right)^{(3/2)}_m ~, 
\nonumber\\
\tilde{P}_{\frac{3}{2},m} &=& \sum_{l=0,2} \alpha_l \, 
\left( a^{\dagger}_{\frac{3}{2}} \times \tilde{b}_l \right)^{(3/2)}_m ~, 
\label{psusy}
\ea
with $\tilde{b}_{l,m}=(-1)^{l-m}b_{l,-m}$~. 
Here we have introduced the notation $b^{\dagger}_0 = s^{\dagger}$ and 
$b^{\dagger}_2 = d^{\dagger}$. The operators of Eq.~(\ref{psusy}) are 
tensor operators under $Spin(5)$ and $Spin(3)$, but not under $Spin(6)$. 

Whereas the fermion creation and annihilation operators of Eqs.~(\ref{pa1}) 
and (\ref{pa2}) transform as tensor operators, the transformation properties 
of the boson creation and annihilation operators depend on the realization 
of the $SO(6)$ algebra. Group theoretically, the standard realization 
of the orthogonal subalgebra is in terms of the antisymmetric generators 
\ba
\bar{Q}^{(2)}_{\mu} &=& 
-i(s^{\dagger} \times \tilde{d} - d^{\dagger} \times \tilde{s})^{(2)}_{\mu } ~,
\nonumber\\
G^{(\lambda)}_{\mu} &=& (d^{\dagger} \times \tilde{d})^{(\lambda)}_{\mu} ~. 
\hspace{3cm} (\lambda=1,3) 
\label{gen1}
\ea
The transformation properties of the boson creation and annihilation 
operators under $SO(6)$ can be investigated by considering the rotation 
\ba
{\cal R}(\theta) &=& \exp \left( -i \theta \, \bar{Q}^{(2)}_0 \right) ~. 
\ea
Since ${\cal R}(\theta)$ transforms the creation and annihilation 
operators in the same way
\ba
{\cal R}(\theta) \left( \begin{array}{c} s^{\dagger} \\ d^{\dagger}_0  
\end{array} \right) {\cal R}^{-1}(\theta) &=& \left( \begin{array}{rr} 
\cos \theta & \sin \theta \\ -\sin \theta & \cos \theta \end{array} \right) 
\left( \begin{array}{c} s^{\dagger} \\ d^{\dagger}_0 \end{array} \right) ~,
\nonumber\\
{\cal R}(\theta) \left( \begin{array}{c} \tilde{s} \\ \tilde{d}_0 
\end{array} \right) {\cal R}^{-1}(\theta) &=& \left( \begin{array}{rr} 
\cos \theta & \sin \theta \\ -\sin \theta & \cos \theta \end{array} \right) 
\left( \begin{array}{c} \tilde{s} \\ \tilde{d}_0 \end{array} \right) ~, 
\label{so6bar}
\ea
the annihilation operators have the same 
tensorial character as the creation operators 
\ba
b^{\dagger}_{2\tau,m} ~, \; \tilde{b}_{2\tau,m} &:& 
T^{(1,0,0)}_{(\tau,0),2\tau,m} ~. \hspace{2cm} (\tau=0,1) 
\ea 
The $SO(6)$ isoscalar factors which are relevant for the $Spin(6)$ limit 
can be derived in the standard way by evaluating the matrix elements of 
the boson-fermion quadrupole-quadrupole interaction \cite{spin6}, provided 
that one uses the antisymmetric quadrupole operator of Eq.~(\ref{gen1}). 
The matrix elements of the latter were derived in \cite{phases}. It is 
important to note that the $SO(6)$ isoscalar factors are related, but not 
equivalent to the expansion coefficients $\xi$ of \cite{spin6,serdar}
\ba
\left\langle \begin{array}{cc|c} 
(N,0,0) & (\frac{1}{2},\frac{1}{2},\frac{1}{2}) 
& (N\pm\frac{1}{2},\frac{1}{2},\mp\frac{1}{2}) \\
(\tau,0),L & (\frac{1}{2},\frac{1}{2}),\frac{3}{2} & 
(\tau+\frac{1}{2},\frac{1}{2}),J \end{array} \right\rangle 
&=&  -i \, \xi^{N,\tau,L}_{N\pm\frac{1}{2},\tau+\frac{1}{2},J} ~,
\nonumber\\
\left\langle \begin{array}{cc|c} 
(N,0,0) & (\frac{1}{2},\frac{1}{2},\frac{1}{2}) 
& (N\pm\frac{1}{2},\frac{1}{2},\mp\frac{1}{2}) \\
(\tau,0),L & (\frac{1}{2},\frac{1}{2}),\frac{3}{2} & 
(\tau-\frac{1}{2},\frac{1}{2}),J \end{array} \right\rangle 
&=& \xi^{N,\tau,L}_{N\pm\frac{1}{2},\tau-\frac{1}{2},J} ~.
\ea
With these isoscalar factors we can construct the single-particle transfer 
operators that connect different members of the same supermultiplet according 
to their tensorial character. We find 
\ba
T^{(\sigma_{1},\sigma_{2},\sigma_{3})}_{(\frac{1}{2},\frac{1}{2}),
\frac{3}{2},m} &=& \sum_{\tau=0,1} \left\langle \begin{array}{cc|c} 
(1,0,0) & (\frac{1}{2},\frac{1}{2},\frac{1}{2}) 
& (\sigma_{1},\sigma_{2},\sigma_{3}) \\
(\tau,0),2\tau & (\frac{1}{2},\frac{1}{2}),\frac{3}{2} & 
(\frac{1}{2},\frac{1}{2}),\frac{3}{2} \end{array} \right\rangle 
\left( \tilde{b}_{2\tau} \times a^{\dagger}_{\frac{3}{2}} 
\right)^{(\frac{3}{2})}_m ~. 
\ea
For the two allowed representations $(\sigma_{1},\sigma_{2},\sigma_{3})$ 
of $Spin(6)$ this gives 
\ba
T^{(\frac{1}{2},\frac{1}{2},-\frac{1}{2})}
_{(\frac{1}{2},\frac{1}{2}),\frac{3}{2},m} &=& -i\sqrt{\frac{1}{6}} 
\left( \tilde{s} \times a^{\dagger}_{\frac{3}{2}} \right)^{(\frac{3}{2})}_m 
+\sqrt{\frac{5}{6}} \left( \tilde{d} \times a^{\dagger}_{\frac{3}{2}} 
\right)^{(\frac{3}{2})}_m ~, 
\nonumber\\ 
T^{(\frac{3}{2},\frac{1}{2},\frac{1}{2})}
_{(\frac{1}{2},\frac{1}{2}),\frac{3}{2},m} &=& i\sqrt{\frac{5}{6}} 
\left( \tilde{s} \times a^{\dagger}_{\frac{3}{2}} \right)^{(\frac{3}{2})}_m 
+\sqrt{\frac{1}{6}} \left( \tilde{d} \times a^{\dagger}_{\frac{3}{2}} 
\right)^{(\frac{3}{2})}_m ~. 
\label{topbar}
\ea

However, the realization commonly used in the IBM is in terms of the 
symmetric quadrupole operator
\ba
Q^{(2)}_{\mu} &=& 
(s^{\dagger} \times \tilde{d} + d^{\dagger} \times \tilde{s})^{(2)}_{\mu } ~,
\nonumber\\
G^{(\lambda)}_{\mu} &=& (d^{\dagger} \times \tilde{d})^{(\lambda)}_{\mu} ~. 
\hspace{3cm} (\lambda=1,3) 
\label{gen2}
\ea
Both realizations, Eqs.~(\ref{gen1}) and (\ref{gen2}), lead to identical 
energy spectra, but the corresponding wave functions differ in relative 
phases. The transformation properties 
of the boson creation and annihilation operators under this realization 
of $SO(6)$ can be investigated by considering the rotation 
\ba
{\cal R}(\theta) &=& \exp \left( -i \theta \, Q^{(2)}_0 \right) ~. 
\ea
In this case, we find the transformation properties 
\ba
{\cal R}(\theta) \left( \begin{array}{c} i \, s^{\dagger} \\ d^{\dagger}_0  
\end{array} \right) {\cal R}^{-1}(\theta) &=& \left( \begin{array}{rr} 
\cos \theta & \sin \theta \\ -\sin \theta & \cos \theta \end{array} \right) 
\left( \begin{array}{c} i \, s^{\dagger} \\ d^{\dagger}_0 \end{array} 
\right) ~,
\nonumber\\
{\cal R}(\theta) \left( \begin{array}{c} -i \, \tilde{s} \\ \tilde{d}_0 
\end{array} \right) {\cal R}^{-1}(\theta) &=& \left( \begin{array}{rr} 
\cos \theta & \sin \theta \\ -\sin \theta & \cos \theta \end{array} \right) 
\left( \begin{array}{c} -i \, \tilde{s} \\ \tilde{d}_0 \end{array} \right) ~, 
\label{so6}
\ea
The transformation properties of Eq.~(\ref{so6}) can thus be expressed 
as the $SO(6)$ rotations of Eq.~(\ref{so6bar}) by carrying out the canonical 
transformation 
\ba
\tilde{s} &\rightarrow& i \, \tilde{s} ~,
\nonumber\\
s^{\dagger} &\rightarrow& -i s^{\dagger} ~,
\nonumber\\
\tilde{d}_m &\rightarrow& \tilde{d}_m ~, 
\nonumber\\
d^{\dagger}_m &\rightarrow& d^{\dagger}_m ~. 
\label{ctr}
\ea
This has important consequences for the transfer operators. The tensor 
operators for single-particle transfer for the realization of $Spin(6)$ 
with the symmetric boson quadrupole operator and the antisymmetric fermion 
quadrupole operator \cite{spin6} can now be obtained by applying the inverse 
of the canonical transformation of Eq.~(\ref{ctr}) to the tensor operators 
of Eq.~(\ref{topbar}). Accordingly we find
\ba
T^{(\frac{1}{2},\frac{1}{2},-\frac{1}{2})}
_{(\frac{1}{2},\frac{1}{2}),\frac{3}{2},m} &=& -\sqrt{\frac{1}{6}} 
\left( \tilde{s} \times a^{\dagger}_{\frac{3}{2}} \right)^{(\frac{3}{2})}_m 
+\sqrt{\frac{5}{6}} \left( \tilde{d} \times a^{\dagger}_{\frac{3}{2}} 
\right)^{(\frac{3}{2})}_m ~, 
\nonumber\\ 
T^{(\frac{3}{2},\frac{1}{2},\frac{1}{2})}
_{(\frac{1}{2},\frac{1}{2}),\frac{3}{2},m} &=& \sqrt{\frac{5}{6}} 
\left( \tilde{s} \times a^{\dagger}_{\frac{3}{2}} \right)^{(\frac{3}{2})}_m 
+\sqrt{\frac{1}{6}} \left( \tilde{d} \times a^{\dagger}_{\frac{3}{2}} 
\right)^{(\frac{3}{2})}_m ~, 
\label{top1} 
\ea
and for the inverse process 
\ba
T^{(\frac{1}{2},\frac{1}{2},\frac{1}{2})}
_{(\frac{1}{2},\frac{1}{2}),\frac{3}{2},m} &=& -\sqrt{\frac{1}{6}} 
\left( s^{\dagger} \times \tilde{a}_{\frac{3}{2}} \right)^{(\frac{3}{2})}_m 
+\sqrt{\frac{5}{6}} \left( d^{\dagger} \times \tilde{a}_{\frac{3}{2}}
\right)^{(\frac{3}{2})}_m ~, 
\nonumber\\ 
T^{(\frac{3}{2},\frac{1}{2},-\frac{1}{2})}
_{(\frac{1}{2},\frac{1}{2}),\frac{3}{2},m} &=& \sqrt{\frac{5}{6}} 
\left( s^{\dagger} \times \tilde{a}_{\frac{3}{2}} \right)^{(\frac{3}{2})}_m 
+\sqrt{\frac{1}{6}} \left( d^{\dagger} \times \tilde{a}_{\frac{3}{2}}
\right)^{(\frac{3}{2})}_m ~. 
\label{top2}
\ea

We note, as is well known, that the annihilation operator $d_m$ by itself 
is not a spherical tensor operator. This led to the introduction of the 
`tilde' operator $\tilde{d}_m=(-1)^m d_{-m}$, which does have the same 
tensor properties as the creation operator $d^{\dagger}_m$. In the present 
case, we have shown that with the usual realization of the $SO(6)$ 
algebra used in the IBM, the boson creation and annihilation operators by 
themselves do not form a $SO(6)$ tensor, but rather 
\ba
\left( \begin{array}{r}  i \, s^{\dagger} \\ d^{\dagger}_m \end{array} 
\right) ~, \left( \begin{array}{r} -i \, \tilde{s} \\ \tilde{d}_m 
\end{array} \right) &:& \left( \begin{array}{c} 
T^{(1,0,0)}_{(0,0),0,0} \\ T^{(1,0,0)}_{(1,0),2,m} \end{array} \right) ~. 
\ea

The use of tensor operators to describe single-particle transfer 
reactions in the supersymmetry scheme has the advantage of giving rise to 
selection rules and closed expressions for the spectroscopic 
factors. Fig.~\ref{spec1} shows the allowed transitions for the 
transfer operators of Eq.~(\ref{top1}) that describe the 
single-particle transfer from the ground state $|(N+1,0,0),(0,0),0\rangle$ 
of the even-even nucleus to the odd-even nucleus belonging to the same 
supermultiplet $[N+1\}$.  
Both operators have the same transformation character 
under $Spin(5)$ and $Spin(3)$, and therefore can only excite states with 
$(\tau_1,\tau_2)=(\frac{1}{2},\frac{1}{2})$ and $J=\frac{3}{2}$. The operators 
differ in their $Spin(6)$ selection rules. Whereas the first operator 
can only excite the ground state band of the odd-even nucleus with 
$(\sigma_1,\sigma_2,\sigma_3)=(N+\frac{1}{2},\frac{1}{2},\frac{1}{2})$, 
the second one allows a transition to an excited band as well 
$(N\pm\frac{1}{2},\frac{1}{2},\pm\frac{1}{2})$. 

Fig.~\ref{spec2} shows the allowed transitions for the inverse process, 
i.e. the single-particle transfer from the ground state 
$|(N+\frac{1}{2},\frac{1}{2},\frac{1}{2}),(\frac{1}{2},\frac{1}{2}),
\frac{3}{2}\rangle$ of the odd-even nucleus to the even-even nucleus 
by the transfer operators of Eq.~(\ref{top2}). Both operators satisfy 
the same selection rules under $Spin(5)$ and $Spin(3)$, which permit the 
excitation of states with $(\tau_1,\tau_2),J=(0,0),0$ or $(1,0),2$. 
According to the $Spin(6)$ selection rules, the first operator 
can only excite the ground state band of the even-even nucleus with 
$(\sigma_1,\sigma_2,\sigma_3)=(N+1,0,0)$, whereas the second one 
allows a transition to an excited band as well $(N \pm 1,0,0)$. 

In Table~\ref{me}, we present the corresponding matrix elements of the 
transfer operators of Eqs.~(\ref{top1}) and (\ref{top2}). 
An analysis of spectroscopic factors for proton pick-up and stripping 
reactions between the pair of nuclei $^{192}$Os and $^{193}$Ir shows that 
the operators $T^{(\frac{1}{2},\frac{1}{2},\mp \frac{1}{2})}$ of 
Eqs.~(\ref{top1}) and (\ref{top2}) decribe the 
main features of the data \cite{spin6}. For the SUSY partners $^{194}$Pt 
and $^{195}$Au, there is to the best of our knowledge only data available 
for the proton stripping reactions $^{194}$Pt$(\alpha,t)^{195}$Au and 
$^{194}$Pt$(^{3}$He$,d)^{195}$Au \cite{munger}. The $J=3/2$ ground state 
of $^{195}$Au is excited strongly with $C^2S=0.175$, whereas the first 
excited $J=3/2$ state is excited weakly with $C^2S=0.019$. In the SUSY 
scheme, the latter state is assigned as a member of the ground 
state band with $(\tau_1,\tau_2)=(5/2,1/2)$. Therefore the 
one proton transfer to this state is forbidden by the $Spin(5)$  
selection rule of the tensor operators of Eq.~(\ref{top1}). The relatively 
small strength to excited $J=3/2$ states suggests that also in this case, 
the operator $T^{(\frac{1}{2},\frac{1}{2},-\frac{1}{2})}$ of 
Eqs.~(\ref{top1}) can be used to describe the data. 

Finally, we note the relevance of tensor operators in the study of 
supersymmetric multiphonon structures \cite{kim}, as well as in the 
construction of supersymmetric `ladder' operators that convert the 
eigenstates of an even-even nucleus into the eigenstates of the odd-even 
nucleus \cite{jolos}. 

\section{Summary and conclusions}

The recent measurements of the spectroscopic properties of the odd-odd 
nucleus $^{196}$Au have rekindled the interest in nuclear supersymmetry. 
The available data on the spectroscopy of the quartet of nuclei 
$^{194}$Pt, $^{195}$Au, $^{195}$Pt and $^{196}$Au can, to a good 
approximation, be described in terms of the $U(6/4) \otimes U(6/12)$ 
supersymmetry. However, there is a still another important set of experiments 
which can further test the predictions of the supersymmetry 
scheme. These involve single-particle transfer reactions between nuclei 
belonging to the same supermultiplet, in particular between the even-odd 
and odd-odd members of the supersymmetric quartet. 
Theoretically, these transfers are described by the supersymmetric 
generators which change a boson into a fermion, or vice versa. 

We investigated in detail the example of proton transfer between the SUSY 
partners $^{194}$Pt and $^{195}$Au in $U(6/4)$, and pointed out that 
special care should be taken in the theoretical analysis because of phase 
ambiguities in the $SO(6)$ algebra. The latter arise because in 
the usual realization of the $Spin(6)$ algebra the symmetric boson 
quadrupole operator is combined with the antisymmetric fermion 
quadrupole operator. We have shown explicitly how in this case the transfer 
operators can be classified according their tensorial character under 
$Spin(6) \supset Spin(5) \supset Spin(3)$. This makes it possible to study 
the corresponding selection rules and derive their matrix elements. 

The problem of the phase ambiguities arises especially for coupled 
systems, such as the neutron-proton IBM, the IBFM, the neutron-proton 
IBFM, and especially the extension to odd-odd nuclei. As an example, 
the odd-odd nucleus $^{196}$Au is described as a system with four  
neutron bosons, one proton boson, one neutron and one proton. 
Moreover, the phase ambiguities not only occur in the $SO(6)$ limit 
of the IBM, but also in the $SU(3)$ limit \cite{smirnov}. For the 
derivation of selection rules and the evaluation of spectroscopic factors 
and transfer intensities, it is crucial to use consistent conventions in 
the construction of the appropiate tensor operators for electromagnetic 
transitions and one-particles transfer reactions. Further work in this 
direction is in progress \cite{work}. To conclude, we emphasize the need 
for new experiments taking advantage of the new experimental capabilities 
\cite{metz,pt195,au196} and suggest that particular attention should be 
paid to one-nucleon transfer reactions between the SUSY partners $^{194}$Pt, 
$^{195}$Au, $^{195}$Pt and $^{196}$Au, since such experiments provide the 
most accurate tests of nuclear supersymmetry. 

\section*{Acknowledgements}

This work was supported in part by CONACyT under projects 
32416-E and 32397-E, by DPAGA-UNAM under project IN106400 
and by DGICYT under project PB98-1111. One of us (J.B.) is 
very grateful to his hosts at the ICN-UNAM.

\clearpage

\begin{table}
\centering
\caption[]{Matrix elements of one-proton transfer operators in the $Spin(6)$ 
limit for transitions (a) from even-even to odd-even and (b) from odd-even 
to even-even. (a) The ground state of the even-even nucleus has 
$\mid i \rangle = \mid [N+1\},[N+1],\{0\},(N+1,0,0),(0,0),0 \rangle$, 
and (b) the ground state of the odd-even nucleus 
$\mid i \rangle = \mid [N+1\},[N],\{1\},(N+\frac{1}{2},\frac{1}{2},
\frac{1}{2}),(\frac{1}{2},\frac{1}{2}),\frac{3}{2} \rangle$. 
For the final states we only show the labels $\langle f \mid = 
\langle (\sigma_1,\sigma_2,\sigma_3),(\tau_1,\tau_2),J \mid$.}
\label{me}
\vspace{15pt}
\begin{tabular}{lcc}
& & \\
(a) $\hspace{1cm} \langle f \mid$  
& $\langle f || \, T^{(\frac{1}{2},\frac{1}{2},-\frac{1}{2})}
_{(\frac{1}{2},\frac{1}{2}),\frac{3}{2}} \, || i \rangle$ 
& $\langle f || \, T^{(\frac{3}{2},\frac{1}{2},\frac{1}{2})}
_{(\frac{1}{2},\frac{1}{2}),\frac{3}{2}} \, || i \rangle$ \\
& & \\
\hline
& & \\
$\langle (N+\frac{1}{2},\frac{1}{2},\frac{1}{2}),(\frac{1}{2},\frac{1}{2}),
\frac{3}{2} \mid$ & $-\sqrt{\frac{2(N+1)}{3}}$ 
& $- \frac{N+5}{N+2} \sqrt{\frac{8(N+1)}{15}}$ \\
& & \\
$\langle (N-\frac{1}{2},\frac{1}{2},-\frac{1}{2}),(\frac{1}{2},\frac{1}{2}),
\frac{3}{2} \mid$ & 0 
& $\frac{\sqrt{6N(N+1)(N+4)/5}}{N+2}$ \\
& & \\
\hline
& & \\
(b) $\hspace{1cm} \langle f \mid$  
& $\langle f || \, T^{(\frac{1}{2},\frac{1}{2},\frac{1}{2})}
_{(\frac{1}{2},\frac{1}{2}),\frac{3}{2}} \, || i \rangle$ 
& $\langle f || \, T^{(\frac{3}{2},\frac{1}{2},-\frac{1}{2})}
_{(\frac{1}{2},\frac{1}{2}),\frac{3}{2}} \, || i \rangle$ \\
& & \\
\hline
& & \\
$\langle (N+1,0,0),(0,0),0 \mid$ & $-\sqrt{\frac{2(N+1)}{3}}$ 
& $-\frac{N+5}{N+2} \sqrt{\frac{8(N+1)}{15}}$ \\
& & \\
$\langle (N+1,0,0),(1,0),2 \mid$ & $\sqrt{\frac{2(N+5)}{3}}$ 
& $\frac{N-1}{N+2} \sqrt{\frac{8(N+5)}{15}}$ \\
& & \\
$\langle (N-1,0,0),(0,0),0 \mid$ & 0 
& $-\frac{1}{N+2} \sqrt{\frac{6N(N+3)(N+4)}{5(N+1)}}$ \\
& & \\
$\langle (N-1,0,0),(1,0),2 \mid$ & 0 
& $ \frac{1}{N+2} \sqrt{\frac{6(N-1)N(N+4)}{5(N+1)}}$\\
& & \\
\end{tabular}
\end{table}

\clearpage

\clearpage

\begin{figure}
\centering
\setlength{\unitlength}{1.0pt}
\begin{picture}(400,200)(0,0)
\thicklines
\put ( 60, 60) {\line(1,0){60}}
\put ( 60,180) {\line(1,0){60}}
\put (280, 60) {\line(1,0){60}}
\put (260, 60) {\vector(-1, 0){120}}
\put (260, 60) {\vector(-1, 1){120}}
\put ( 60, 40) {$(N+\frac{1}{2},\frac{1}{2},\frac{1}{2})$}
\put ( 10, 55) {$(\frac{1}{2},\frac{1}{2}),\frac{3}{2}$}
\put ( 60,160) {$(N-\frac{1}{2},\frac{1}{2},-\frac{1}{2})$}
\put ( 10,175) {$(\frac{1}{2},\frac{1}{2}),\frac{3}{2}$}
\put (280, 40) {$(N+1,0,0)$}
\put (345, 55) {$(0,0),0$}
\end{picture}
\vspace{1cm}
\caption[]{Allowed single-particle transfer reactions for 
transitions from even-even to odd-even nuclei in $U(6/4)$.}
\label{spec1}
\end{figure}

\begin{figure}
\centering
\setlength{\unitlength}{1.0pt}
\begin{picture}(400,200)(0,0)
\thicklines
\put ( 60, 60) {\line(1,0){60}}
\put (280, 60) {\line(1,0){60}}
\put (280, 90) {\line(1,0){60}}
\put (280,150) {\line(1,0){60}}
\put (280,180) {\line(1,0){60}}
\put (140, 60) {\vector( 1, 0){120}}
\put (140, 60) {\vector( 4, 1){120}}
\put (140, 60) {\vector( 4, 3){120}}
\put (140, 60) {\vector( 1, 1){120}}
\put ( 60, 40) {$(N+\frac{1}{2},\frac{1}{2},\frac{1}{2})$}
\put ( 10, 55) {$(\frac{1}{2},\frac{1}{2}),\frac{3}{2}$}
\put (280, 40) {$(N+1,0,0)$}
\put (345, 55) {$(0,0),0$}
\put (345, 85) {$(1,0),2$}
\put (280,130) {$(N-1,0,0)$}
\put (345,145) {$(0,0),0$}
\put (345,175) {$(1,0),2$}
\end{picture}
\vspace{1cm}
\caption[]{Allowed single-particle transfer reactions for 
transitions from odd-even to even-even nuclei in $U(6/4)$.}
\label{spec2}
\end{figure}


\begin{thebibliography}{99}

\bibitem{metz}
A. Metz, J. Jolie, G. Graw, R. Hertenberger, J. Gr\"oger, 
C. G\"unther, N. Warr and Y. Eisermann, 
Phys. Rev. Lett. {\bf 83}, 1542 (1999). 

\bibitem{pt195}
A. Metz, Y. Eisermann, A. Gollwitzer, R. Hertenberger, B.D. Valnion, 
G. Graw and J. Jolie, 
Phys. Rev. C {\bf 61}, 064313 (2000).

\bibitem{au196}
J. Gr\"oger, J. Jolie, R. Kr\"ucken, C.W. Beausang, M. Caprio, R.F. Casten, 
J. Cederkall, J.R. Cooper, F. Corminboeuf, L. Genilloud, G. Graw, 
C. G\"unther, M. de Huu, A.I. Levon, A. Metz, J.R. Novak, N. Warr and 
T. Wendel, 
Phys. Rev. C {\bf 62}, 064304 (2000).

\bibitem{baha}
A.B. Balantekin, I. Bars, R. Bijker and F. Iachello, 
Phys. Rev. C {\bf 27}, 1761 (1983). 

\bibitem{sun}
H.Z. Sun, A. Frank and P. van Isacker, 
Phys. Rev. C {\bf 27}, 2430 (1983). 

\bibitem{magic}
P. Van Isacker, J. Jolie, K. Heyde and A. Frank,
Phys. Rev. Lett. {\bf 54}, 653 (1985).

\bibitem{mauthofer}
A. Mauthofer, K. Stelzer, J. Gerl, Th.W. Elze, Th. Happ, G. Eckert, 
T. Faestermann, A. Frank and P. van Isacker, 
Phys. Rev. C {\bf 34}, 1958 (1986). 

\bibitem{jolie}
J. Jolie, U. Mayerhofer, T. von Egidy, H. Hiller, J. Klora, H. Lindner 
and H. Trieb, 
Phys. Rev. C {\bf 43}, R16 (1991). 

\bibitem{rotbard}
G. Rotbard, G. Berrier, M. Vergnes, S. Fortier, J. Kalifa, J.M. Maison, 
L. Rosier, J. Vernotte, P. van Isacker and J. Jolie, 
Phys. Rev. C {\bf 47}, 1921 (1993).

\bibitem{IBM}
F. Iachello and A. Arima, 
`The interacting boson model', 
(Cambridge University Press, 1987). 

\bibitem{IBFM}
F. Iachello and P. van Isacker, 
`The interacting boson-fermion model', 
(Cambridge University Press, 1991). 

\bibitem{FI}
F. Iachello,
Phys. Rev. Lett. {\bf 44}, 772 (1980). 

\bibitem{susy}
A.B. Balantekin, I. Bars and F. Iachello, 
Phys. Rev. Lett. {\bf 47}, 19 (1981);\\
A.B. Balantekin, I. Bars and F. Iachello, 
Nucl. Phys. A {\bf 370}, 284 (1981). 

\bibitem{phases}
P. Van Isacker, A. Frank and J. Dukelsky, 
Phys. Rev. C {\bf 31}, 671 (1985). 

\bibitem{spin6}
F. Iachello and S. Kuyucak, 
Ann. Phys. (N.Y.) {\bf 136}, 19 (1981).

\bibitem{olaf}
O. Scholten, 
Prog. Part. Nucl. Phys. {\bf 14}, 189 (1985). 

\bibitem{BI}
R. Bijker and F. Iachello, 
Ann. Phys. (N.Y.) {\bf 161}, 360 (1985). 

\bibitem{serdar}
S. Kuyucak,
`Study of Spinor Symmetries in Nuclear Structure', 
Ph.D. thesis, Yale University (1982). 

\bibitem{munger}
M.L. Munger and R.J. Peterson, 
Nucl. Phys. A {\bf 303}, 199 (1978).

\bibitem{kim}
K.-H. Kim, T. Otsuka, A. Gelberg, P. von Brentano and P. van Isacker, 
Phys. Rev. Lett. {\bf 76}, 3514 (1996). 

\bibitem{jolos}
R.V. Jolos, P. von Brentano, A. Gelberg, K.-H. Kim and T. Otsuka, 
Phys. Lett. B {\bf 430}, 1 (1998). 

\bibitem{smirnov} 
A.M. Shirokov, N.A. Smirnova, Yu.F. Smirnov, O. Casta\~nos and A. Frank, 
Phys. Atom. Nucl. {\bf 63}, 760 (2000).

\bibitem{work}
J. Barea, G. Loyola, R. Bijker and A. Frank, to be published. 

\end{thebibliography}
\end{document}